\documentclass[fleqn,usenatbib]{mnras}
\usepackage{newtxtext,newtxmath}
\usepackage{txfonts}

\usepackage[T1]{fontenc}

\DeclareRobustCommand{\VAN}[3]{#2}
\let\VANthebibliography\thebibliography
\def\thebibliography{\DeclareRobustCommand{\VAN}[3]{##3}\VANthebibliography}

\usepackage{graphicx}	% Including figure files
\usepackage{amsmath}	% Advanced maths commands
\usepackage{amssymb}	% Extra maths symbols
\usepackage{adjustbox}
\usepackage{comment}
\usepackage{tablefootnote}
\usepackage{ulem}

\newcommand{\orcid}[1]{\href{https://orcid.org/#1}{\includegraphics[height=11pt]{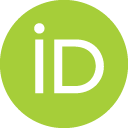}}}

\newcommand{\Feh}{\textit{[Fe/H]$_{Current}$}}
\newcommand{\Afe}{\textit{[$\alpha$/Fe]$_{Current}$}}

\newcommand{\dfeh}{$\Delta$\textit{[Fe/H]}}
\newcommand{\dafe}{$\Delta$\textit{[$\alpha$/Fe]}}

\newcommand{\feh}{\textit{[Fe/H]$_{Birth}$}}
\newcommand{\afe}{\textit{[$\alpha$/Fe]$_{Birth}$}}

\title[SN Ia Progenitor Star Birth Environments]{Tracing back the birth environments of Type Ia supernova progenitor stars: A pilot study based on 44 early-type host galaxies}

\author[Young-Lo Kim et al.]{
	Young-Lo Kim$^{1}$\thanks{E-mail: y.kim9@lancaster.ac.uk}\orcid{0000-0002-1031-0796}, 
	Llu\'is Galbany$^{2, 3}$\orcid{0000-0002-1296-6887},
	Isobel Hook$^{1}$\orcid{0000-0002-2960-978X},
	Yijung Kang$^{4}$\orcid{0000-0002-5261-5803}
\\
$^{1}$Department of Physics, Lancaster University, Lancs LA1 4YB, UK \\
$^{2}$ Institute of SpaceSciences (ICE,CSIC), Campus UAB, Carrer de Can Magrans, s/n, E-08193 Barcelona, Spain \\
$^{3}$ Institut d’Estudis Espacials de Catalunya (IEEC), E-08034 Barcelona, Spain \\
$^{4}$ Kavli Institute for Particle Astrophysics and Cosmology, SLAC National Accelerator Laboratory, Stanford University, 2575 Sand Hill Road, Menlo Park, CA 94025, USA
}

\date{Accepted XXX. Received YYY; in original form ZZZ}

\pubyear{2024}

\begin{document}
\label{firstpage}
\pagerange{\pageref{firstpage}--\pageref{lastpage}}
\maketitle

% Abstract of the paper
\begin{abstract}
The environmental dependence of Type Ia supernova (SN Ia) luminosities is well-established, and efforts are being made to find its origin.
Previous studies typically use the currently-observed status of the host galaxy.
However, given the delay time between the birth of the progenitor star and the SN Ia explosion, the currently-observed status may differ from the birth environment of the SN Ia progenitor star.
In this paper, employing the chemical evolution and accurately determined stellar population properties of 44 early-type host galaxies, we, for the first time, estimate the SN Ia progenitor star birth environment, specifically \feh{} and \afe{}.
We show that \afe{} has a $30.4^{\text{+10.6}}_{-10.1}\%$ wider range than the currently-observed \Afe{}, while the range of \feh{} is not statistically different ($17.9^{\text{+26.0}}_{-27.1}\%$) to that of \Feh{}.
The birth and current environments of [Fe/H] and [$\alpha$/Fe] are sampled from different populations ($p$-values of the Kolmogorov-Smirnov test < 0.01).
We find that light-curve fit parameters are insensitive to \feh{} ($<0.9\sigma$ for the non-zero slope), while a linear trend is observed with HRs at the 2.4$\sigma$ significance level.
With \afe{}, no linear trends ($<1.1\sigma$) are observed.
Interestingly, we find that \afe{} clearly splits the SN Ia sample into two groups: SN Ia exploded in \afe{}-rich or \afe{}-poor environments.
SNe Ia exploded in different \afe{} groups have different weighted-means of light-curve shape parameters: $0.81\pm0.33$ ($2.5\sigma$).
They are thought to be drawn from different populations ($p$-value = 0.01).
Regarding SN Ia colour and HRs, there is no difference ($<1.0\sigma$) in the weighted-means and distribution ($p$-value > 0.27) of each \afe{} group.
\end{abstract}

% Select between one and six entries from the list of approved keywords.
% Don't make up new ones.
\begin{keywords}
supernovae: general -- galaxies: fundamental parameters -- methods: data analysis
\end{keywords}

%%%%%%%%%%%%%%%%%%%%%%%%%%%%%%%%%%%%%%%%%%%%%%%%%%

%%%%%%%%%%%%%%%%% BODY OF PAPER %%%%%%%%%%%%%%%%%%

%%%
%%%
%			Section: Introduction
%%%
%%%
\section{Introduction}
\label{sec:intro}

Applying empirical light-curve corrections, Type Ia supernovae (SNe Ia) play crucial roles in determining the distance of external galaxies \citep{Phillips1993, Guy2007, Jha2007}.
However, the physics of SNe Ia, such as the progenitor stars and the explosion mechanisms, are still not fully understood.
For understanding the progenitor stars, current studies are using host galaxies.
It is well-established that SN Ia luminosities, after light-curve corrections, are correlated with various physical properties of their host galaxies at global \citep[e.g.,][]{Sullivan2010, Childress2013, Pan2014, Kim2019, Smith2020} and local scales \citep[e.g.,][]{Rigault2013, Kim2018, Rigault2020, Kelsey2021, Kim2024}. 

Possible explanations for the origin of the correlations between SN Ia luminosities and host galaxy properties are discussed in intrinsic and/or extrinsic respects.
Intrinsic explanations focus on different properties of the SN Ia progenitor stars, such as age and metallicity.
This is suggested theoretically \citep{Timmes2003, Kasen2009, Childress2014} and empirically based on the host galaxy analysis \citep[e.g., ][]{Sullivan2010, Kang2016, Kim2018, Kim2019, Kang2020, Rigault2020}.
Recently, differences in dust properties in host galaxies have been suggested as the extrinsic explanation \citep{Brout2021}.
In addition, efforts are being made to combine both intrinsic and extrinsic explanations \citep{Kelsey2023, Wiseman2023}.

However, these host galaxy studies and possible explanations for the origin are discussed based on the currently-observed status of host galaxies at the time of SN Ia explosions.
Considering the delay time, the time difference between the birth of the progenitor star and the explosion of SNe Ia \citep{Maoz2012}, this currently-observed status may not represent the birth environments of the SN Ia progenitor stars \citep[see][for a discussion]{Millan-Irigoyen2022}.
Furthermore, even if the progenitor star is identified in a pre-SN explosion image via the archival data search after the SN Ia explosion, the final state of the progenitor star may differ from its state in the birth environment.
This is because the progenitor star evolved while interacting with its environment, such as its companion star in a binary system and dust around it.
Therefore, in order to accurately investigate the origin of the environmental dependence of SN Ia luminosities, and, ultimately, to understand the progenitor star, the birth environment should be adopted.

In the present work, we try to trace back the birth environment of the SN Ia progenitor star, employing the chemical evolution of galaxies, specifically early-type galaxies.
Early-type galaxies, in general, are homogeneous in terms of the stellar population, because their stellar populations are formed through a single burst of star formation followed by passive evolution \citep[e.g.,][]{Thomas2005}.
Thus, their star formation history can be represented by a single Gaussian distribution \citep{Thomas2005, Thomas2010}.
However, it is known that at least 15 percent of nearby giant early-type galaxies show a sign of recent ($\le$ 1 Gyr) star formation \citep[e.g.,][]{Yi2005, Gomes2016a, Gomes2016b}. 
In this situation, if we can select a genuine early-type galaxy (e.g., absorption-dominated spectra from passive environments without recent or ongoing star formation) and a SN Ia is observed in this genuine early-type galaxy, it is most likely that the SN Ia progenitor star formed simultaneously with the formation of the single stellar population.
Therefore, in the case of genuine early-type host galaxies, the galaxy properties when the galaxy is 0 years old (i.e. when most of the stars formed) could represent the birth environment of the SN Ia progenitor star.

Regarding the SN Ia progenitor star birth environment, we will determine [Fe/H] and [$\alpha$/Fe] of the birth environment.
[Fe/H] is iron abundance relative to hydrogen and is commonly used as a tracer of stellar metallicity.
[$\alpha$/Fe] is the enrichment of $\alpha$-process elements (e.g., O, Ne, and Ni) relative to iron, also called the $\alpha$-enhancement.
Since stellar evolution is obviously affected by metallicity (i.e., the abundance of elements heavier than hydrogen and helium), it would be natural to assume that SN Ia luminosity should depend on metallicity.
\citet{Timmes2003} showed that a variation of a factor of three in the initial metallicity, defined in their paper as the CNO + Fe abundances, leads to up to 0.2 mag difference in the peak $V$-band brightness \citep[see also][]{Kasen2009, Hoflich2010}.
However, estimating the progenitor star's metallicity through observation is really challenging.
The present work will determine [Fe/H] and [$\alpha$/Fe] of the SN Ia progenitor star birth environment.
Since the progenitor star formed in this birth environment, the determined [Fe/H] and [$\alpha$/Fe] can be considered the progenitor star's metallicity, which provides more direct information about the progenitor star than the currently-widely used stellar mass of the host galaxy.
We found that both properties are available in \citet{Kang2016, Kang2020}, and hence we employ them for the present work.

In Sec.~\ref{sec:method}, we trace back the progenitor star birth environments for SNe Ia with accurately determined stellar population properties of their early-type host galaxies.
Then, Sec.~\ref{sec:results} presents the difference between currently-observed and birth environments, and the relationships between the birth environment and SN Ia properties.
We discuss and conclude in Sec.~\ref{sec:discussion}.

%%%
%%%
%			SECTION: Method
%%%
%%%
\section{Method}
\label{sec:method}

% FIGURE: Ex. host spectrum
\begin{figure*}
 \centering
  	\includegraphics[width=\textwidth]{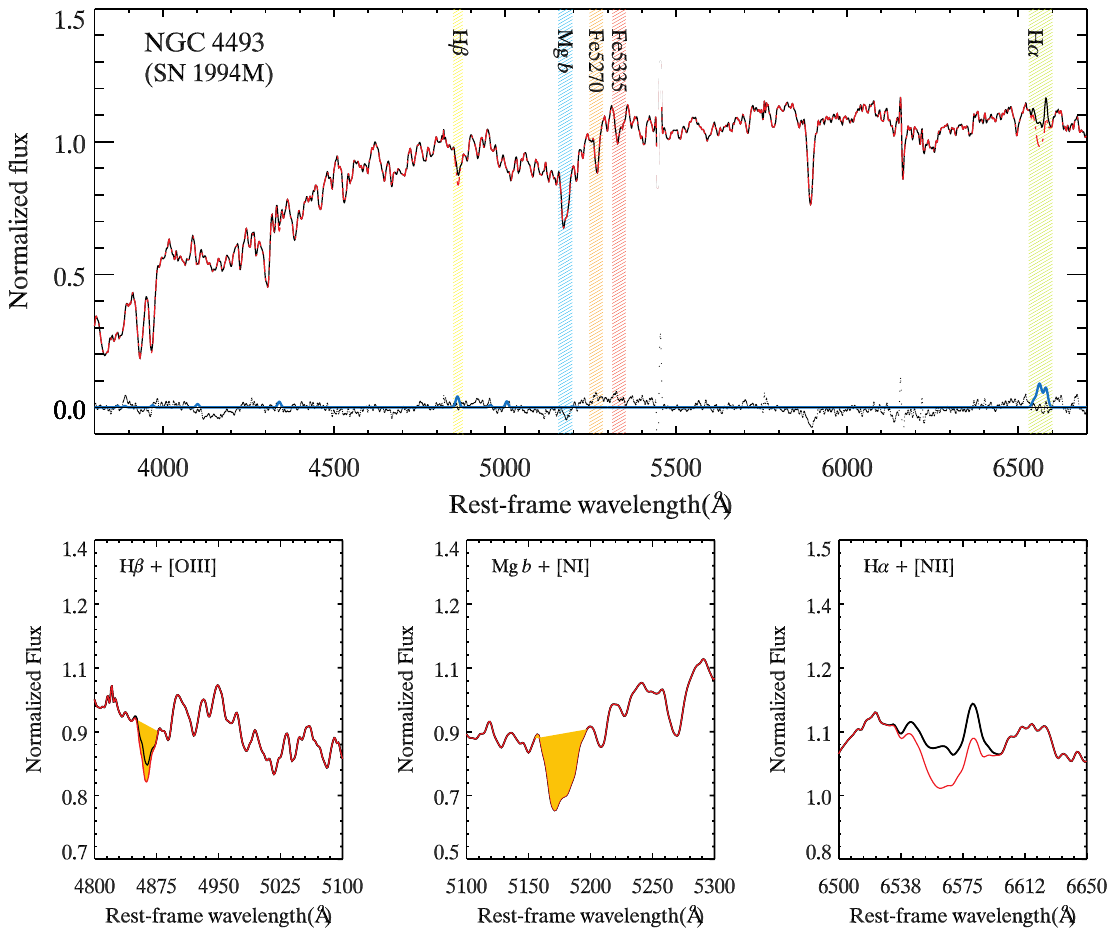}
  \caption{Example of a host galaxy spectrum of NGC 4493.
  		The signal-to-noise ratio of the spectrum is $\sim$148.9.
		The black and red solid lines are fully calibrated spectra in the rest-frame, before and after the emission correction, respectively.
		The upper panel shows the entire wavelength coverage of observation, where the absorption bands for H$\beta$, Mg $b$, Fe5270, Fe5335, and H$\alpha$ are indicated in colour shades.
		The grey line at the bottom of the panel shows the difference between the best-fit model and the spectrum, where the detected emission lines (cyan) are overlapped on the residuals.
		The lower panels show the spectral regions around H$\beta$ + [O III], Mg $b$ + [N I], and H$\alpha$+[N II].
		This figure and the caption (with a minor change) are taken from Fig. 2 of \citet{Kang2016}.
	       }
  \label{fig:ex_sp}
\end{figure*}

To trace back the birth environment of the SN Ia progenitor star, we employ the chemical, specifically [Fe/H] and [$\alpha$/Fe], evolution of galaxies.
Inside one galaxy, [Fe/H] and [$\alpha$/Fe] would increase with the stellar population age (e.g., see \citealt{Dong2018, Saglia2018} for the M31 bulge and \citealt{Williams2017} for the M31 disk observations and stellar population analyses).
For this, we employ correlations of stellar population age with [Fe/H] and [$\alpha$/Fe] calculated by \citet{Walcher2015} to find the relative difference between them:

\begin{equation}
\label{eq:1}
\Delta \left[ Fe \text{/} H \right] \text{= } 0.02 \left(\pm 0.001\right) \times \Delta Age \left(Gyr \right)
\end{equation}

\begin{equation}
\label{eq:2}
\Delta \left[ \alpha \text{/} Fe \right] \text{= } 0.02 \left(\pm 0.001\right) \times \Delta Age \left(Gyr \right)
\end{equation}

In order to utilize these correlations, the currently-observed status of host galaxies, such as stellar population age, metallicity (\Feh{}), and $\alpha$-elements (\Afe{}), is required.
Those properties are taken from \citet{Kang2016, Kang2020}, which determined the properties of 51 early-type host galaxies.
Briefly, \citet{Kang2016, Kang2020} observed early-type host galaxies from the SN Ia catalogue collected by \citet{Kim2019}, covering most of the reported nearby (0.01 < z < 0.08) early-type host galaxies up to that time. 
The galaxy morphological classification is from the NASA Extragalactic Database\footnote{\href{https://ned.ipac.caltech.edu}{https://ned.ipac.caltech.edu}} or the HyperLeda database \citep{Makarov2014}.
Then, using the Yonsei evolutionary population synthesis model \citep{Chung2013}{\footnote{We present in Sec.~\ref{app:other_models} the results with other models employed in \citet{Kang2020}.}, they accurately determined the host galaxy stellar population age and metallicity ([M/Fe]) from absorption line analysis of observed very high-quality spectra (mean signal-to-noise ratio $\sim$ 175; see Fig~\ref{fig:ex_sp}).
However, for using equations~\ref{eq:1} and \ref{eq:2}, we need [Fe/H] and  [$\alpha$/Fe].
Those properties were provided by one of the authors (Y. Kang), who determined them in \citet{Kang2020}.

Among 51 early-type galaxies, we need to select only genuine early-type galaxies.
\citet{Kang2020} provide 11 non-genuine early-type galaxies.
Three galaxies show UV/IR excess, indicating significant ongoing or recent star formation activities, identified from an UV-optical-IR colour-colour diagram.
Eight galaxies younger than 2.5 Gyr are classified as rejuvenated (by recent minor star formation) galaxies based on the result of \citet[see also \citealt{Yi2005}]{Thomas2010}.
Because those non-genuine early-type galaxies are not homogeneous in terms of stellar populations, it would not be appropriate to trace back the birth environments with our methodology, which will adopt a single stellar population model.
Therefore, we do not include them in our analysis, except for four very young ($<$0.1 Gyr) galaxies among rejuvenated galaxies, because their current and birth environments are very similar ($\Delta\left[Fe\text{/}H\right]$ $\sim$ $\Delta\left[\alpha\text{/}Fe\right]$ $\sim$ $0.002$).
In total, we use 44 (out of 51) early-type host galaxies.

Given the age of each host galaxy, we can determine the difference between the currently-observed age and when the galaxy was 0 Gyr old ($\Delta Age$).
Then, from equations~\ref{eq:1} and ~\ref{eq:2}, we can estimate \dfeh{} and \dafe{}.
Finally, [Fe/H] and [$\alpha$/Fe] at the SN Ia progenitor star birth environment (\feh{} and \afe{}, respectively) are determined from

\begin{equation}
\label{eq:3}
\left[ Fe \text{/} H \right]_{Birth} \text{= } \left[ Fe \text{/} H \right]_{Current} - \Delta \left[ Fe \text{/} H \right] 
\end{equation}

\begin{equation}
\label{eq:4}
\left[ \alpha \text{/} Fe \right]_{Birth} \text{= } \left[ \alpha \text{/} Fe \right]_{Current} - \Delta \left[ \alpha \text{/} Fe \right].
\end{equation}

An uncertainty for the birth environment is determined considering the best-fitted value, upper and lower errors of currently-observed host galaxy properties with slope and its errors of Eq.s~\ref{eq:1} or \ref{eq:2} depending on the birth environment.
This gives a total of 27 calculations for each birth environment.
Among them, a value estimated from the best-fitted currently-observed properties and the slope was taken as a representative value of the birth environment, and the difference between this representative value and the maximum (minimum) value was considered an upper (lower) uncertainty.

%%%
%%%
%			SECTION:Results
%%%
%%%
\section{Results}
\label{sec:results}

%%%
%			Subsection: Current vs. Birth, and Isochrones
%%%

% FIGURE: Isochrones
\begin{figure*}
 \centering
  	\includegraphics[width=0.33\textwidth]{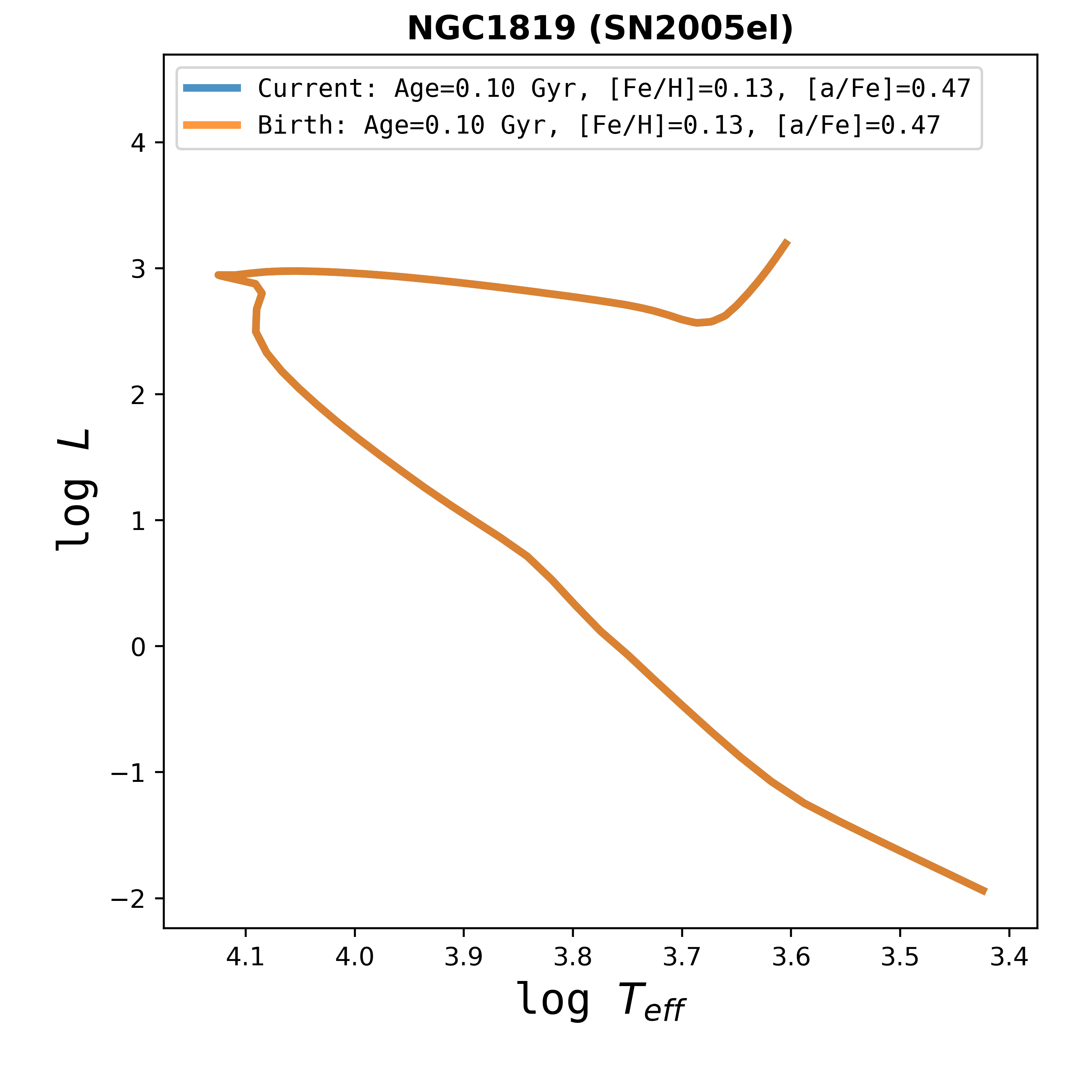}
  	\includegraphics[width=0.33\textwidth]{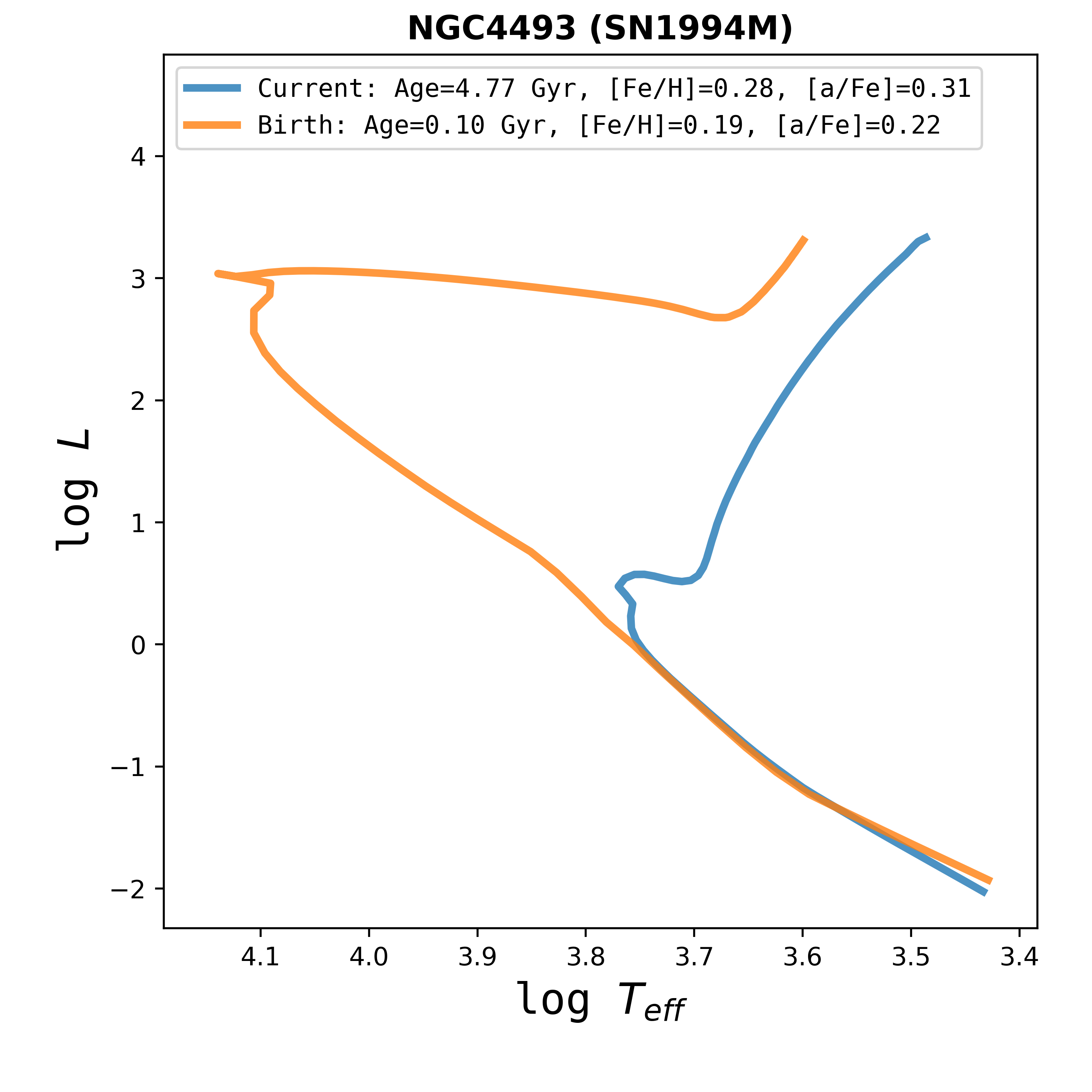}
  	\includegraphics[width=0.33\textwidth]{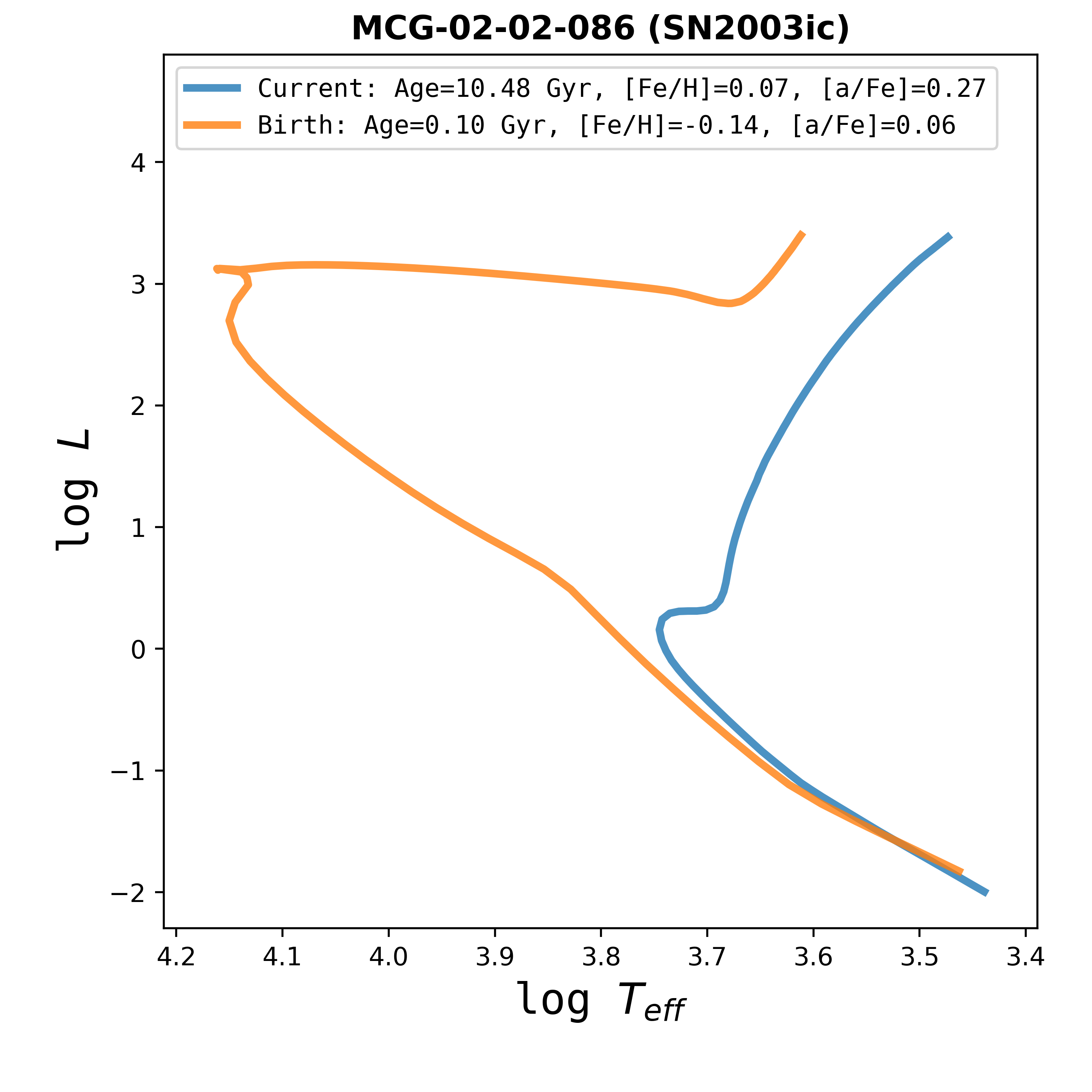}
  \caption{Examples of isochrones for the youngest, median, and oldest host galaxies in our sample.
  	       Blue isochrones indicate the currently-observed status of the host galaxies, while orange isochrones show the SN Ia progenitor star birth environments.
	       We note that for illustration purpose, we use $Age$ = 0.10 Gyr instead of 0.0 Gyr for the birth age, but for the analysis, we use \feh{} and \afe{} at $Age$ = 0.0 Gyr.
	       Because of this, in the youngest, isochrones of currently-observed and birth environments overlap.
	       }
  \label{fig:iso}
\end{figure*}

\subsection{Comparison of currently-observed and SN Ia progenitor star birth environments}
\label{subsec:comp_env}

To visually illustrate the difference between the currently-observed and the SN Ia progenitor birth environments, we construct isochrones based on the galaxy stellar population age, [Fe/H], and [$\alpha$/Fe].
Because a base model of \citet{Kang2016, Kang2020} was constructed from Yonsei-Yale stellar isochrones\footnote{\href{http://cascade.yonsei.ac.kr/~yckim/yyiso.html}{http://cascade.yonsei.ac.kr/~yckim/yyiso.html} }\citep{Kim2002, Demarque2004}, we also employ them to draw isochrones for each host galaxy.
Fig.~\ref{fig:iso} presents examples of host galaxy isochrones we extracted.
Blue isochrones show the currently-observed status of the host galaxy, while orange isochrones represent when the host galaxy is 0 Gyr old, i.e., the SN Ia progenitor star birth environment.
The figure shows how different the two environments are: the older the galaxy, the larger the difference (from the left to the right panel).

% FIGURE: Current vs. Birth [Fe/H] and [a/Fe].
\begin{figure*}
 \centering
  	\includegraphics[width=1.2\columnwidth]{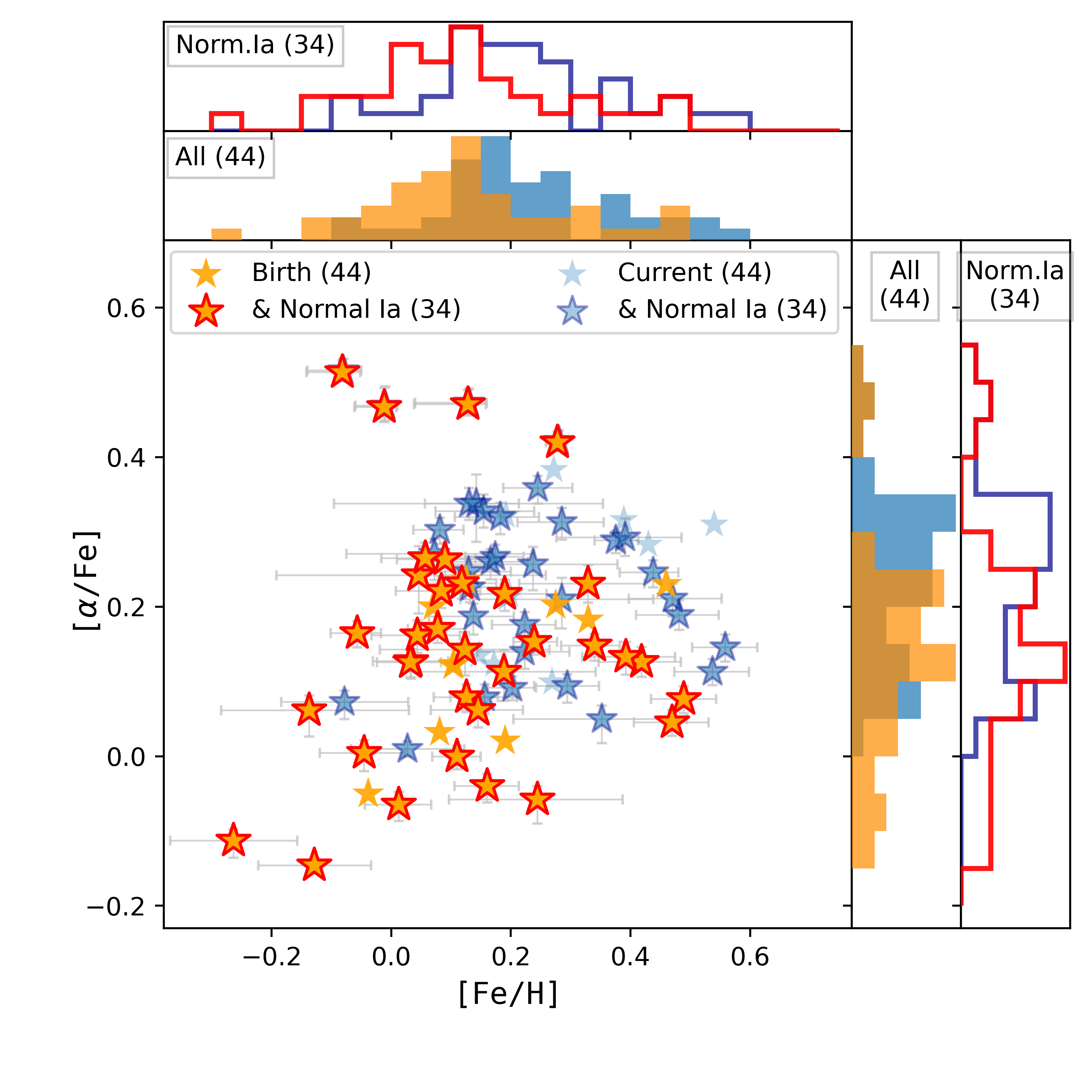}
	\includegraphics[width=0.8\columnwidth]{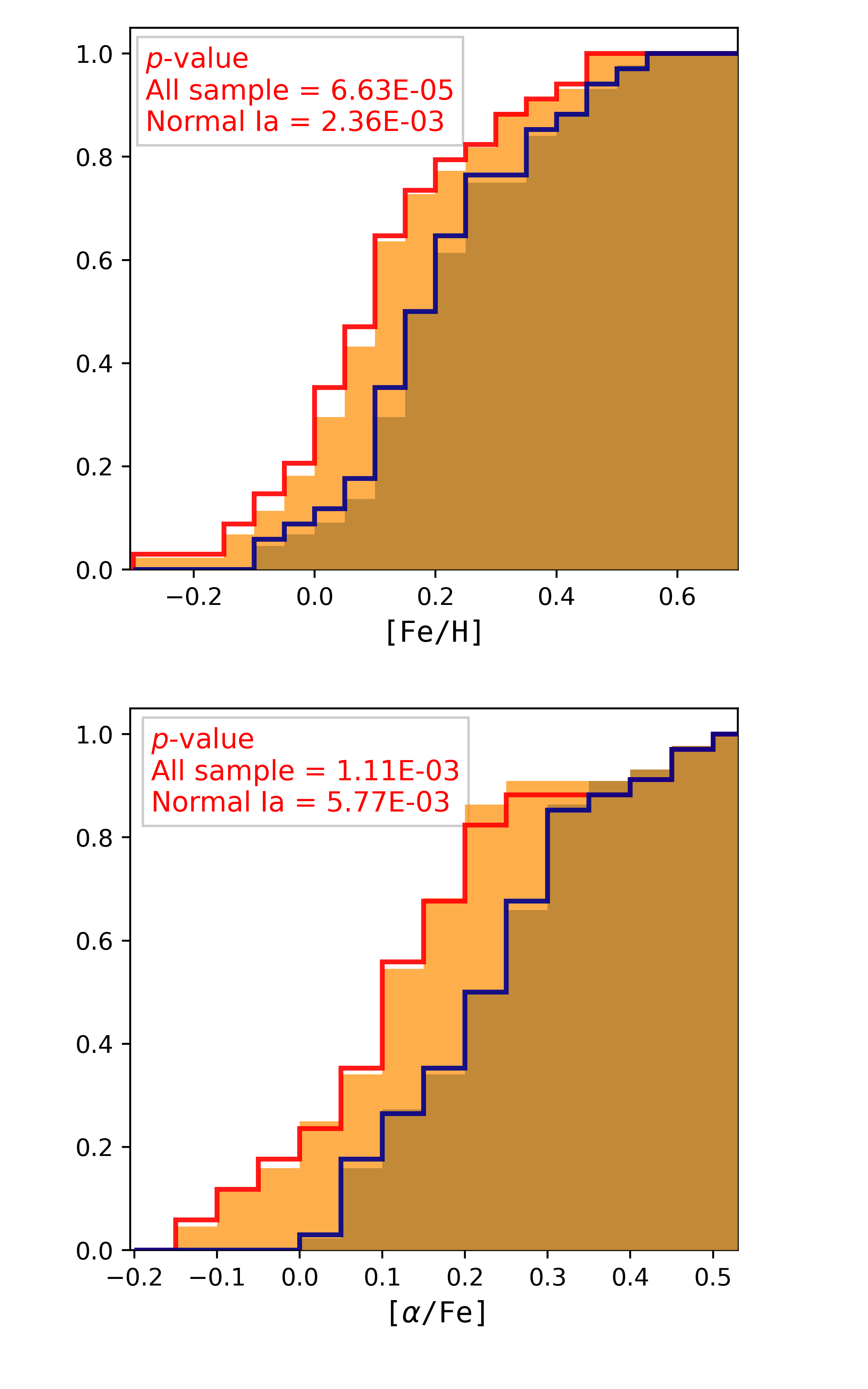}
  \caption{Distributions of SN Ia progenitor star birth (orange) and currently-observed (cyan) [Fe/H] and [$\alpha$/Fe] for each of the 44 host galaxies.
  		The 34 hosts of normal SNe Ia (see Sec.~\ref{subsec:birth_env_vs_salt2} for their selection criteria) are indicated with solid lines around each star-shaped mark.
                The $p$-values of the two-sample Kolmogorov-Smirnov test are presented in the right panels together with cumulative distributions.
                The birth environments for all and normal SN Ia samples have $17.9^{\text{+26.0}}_{-27.1}\%$ and $30.4^{\text{+10.6}}_{-10.1}\%$ wider ranges for [Fe/H] and [$\alpha$/Fe] than the currently-observed environments, respectively.
                The $p$-values in the right panels show that birth and current environments are sampled from populations with different distributions, regardless of all or normal Ia host samples.
                }
  \label{fig:comp_feh_afe}
\end{figure*}

More specifically, we present the difference in [Fe/H] and [$\alpha$/Fe] between the SN Ia progenitor birth and currently-observed environments in Fig.~\ref{fig:comp_feh_afe} for 44 host galaxies and also 34 hosts of normal SNe Ia (see Sec.~\ref{subsec:birth_env_vs_salt2} for their selection criteria).
At first glance, the figure shows that the birth environments are more spread-out than the currently-observed environments.
The birth environments have $17.9^{\text{+26.0}}_{-27.1}\%$ and $30.4^{\text{+10.6}}_{-10.1}\%$ wider ranges for [Fe/H] and [$\alpha$/Fe], respectively, regardless of all or normal Ia host samples.
Due to the wider range, especially in [$\alpha$/Fe], a clear separation in \afe{} is observed, implying distinct groups in the SN Ia progenitor star.

Further, we performed the two-sample Kolmogorov-Smirnov (KS) test to determine if birth and current environments are drawn from the same distribution or not (the right panels of Fig.~\ref{fig:comp_feh_afe}).
The test returned the $p$-values of $6.63\times10^{-5}$ for [Fe/H] and $1.11\times10^{-3}$ for [$\alpha$/Fe] when considering all 44 hosts.
For the 34 normal Ia hosts, $p$-values are $2.36\times10^{-3}$ for [Fe/H] and $5.77\times10^{-3}$ for [$\alpha$/Fe].
This means that birth and current environments are sampled from populations with different distributions, regardless of the host sample.

%%%
%		Subsection: SALT2 parameters vs. Birth Env.
%%%
\subsection{Impact of the SN Ia progenitor star birth environments on the SN observables}
\label{subsec:birth_env_vs_salt2}

% FIGURE: SATL2 vs. birth Envs.
\begin{figure*}
 \centering
  	\includegraphics[scale=0.6]{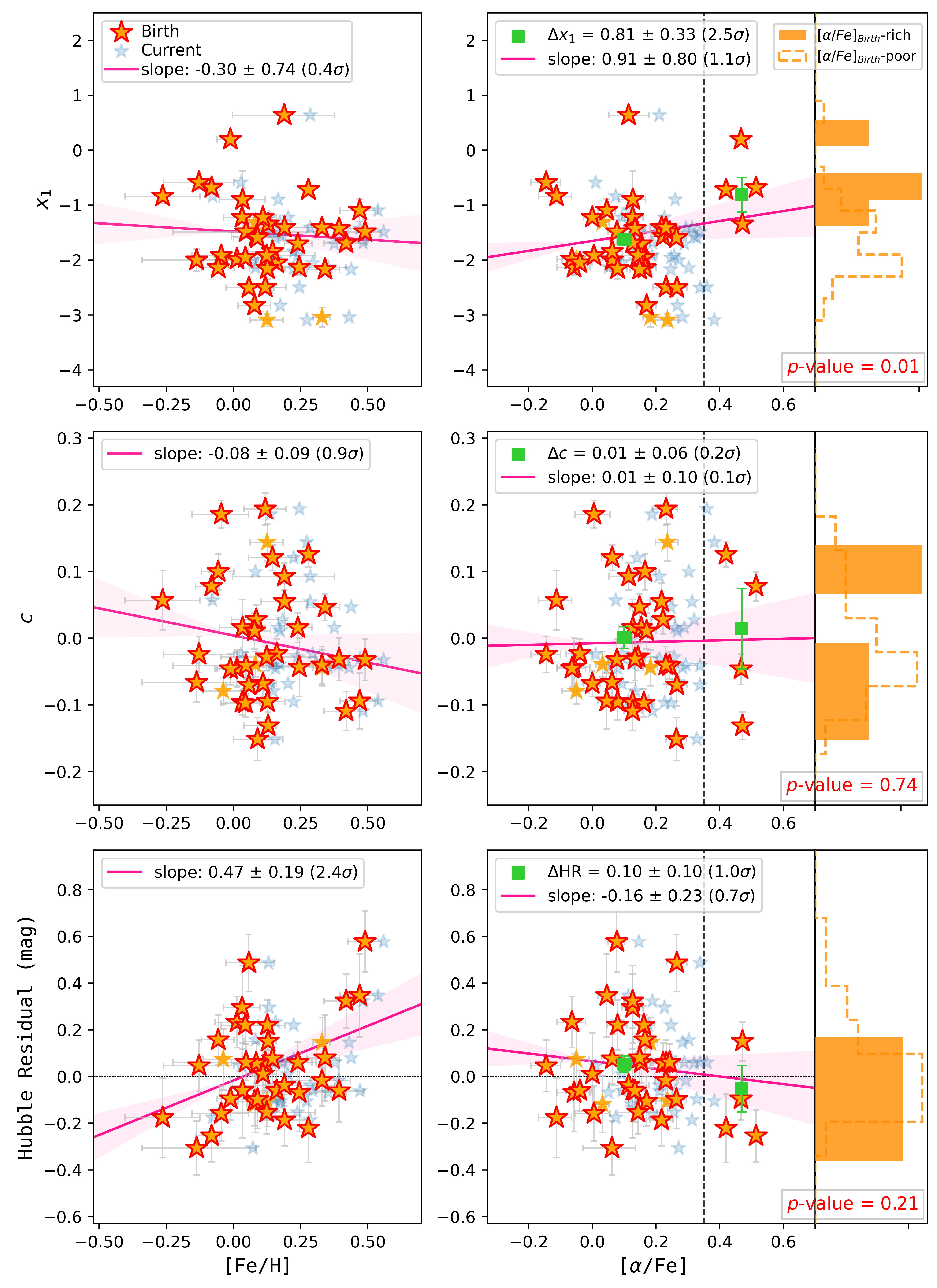}
  \caption{SALT2.4 light-curve fit parameters and Hubble residuals as a function of the SN Ia progenitor birth star (orange stars) and currently-observed (faint blue) environments for 38 SNe Ia in our sample.
  		Among those, four light orange stars are SNe Ia that do not pass the typical cuts applied when making a cosmological sample, such as $|x_{1}| < 3$ and $E\left(B - V\right)_{MW} < 0.15$.
                We do not include those when calculating statistics.
                Pink lines show the slope measurements from a linear least-square regression fit, and pink shaded regions present the standard error of the estimated slope.
                Green squares in the right panel represent the weighted-means of \afe{}and each SN Ia parameter in different [$\alpha$/Fe] environments, split at \afe{} = 0.35 (black dashed lines; cf. \Afe{} = 0.40 based on the \afe{} split value).
                Their legends show the difference in each parameter in different \afe{} environments.
                The $p$-values of the two-sample Kolmogorov-Smirnov test and the distribution for SNe Ia in different \afe{} environments (orange histograms) are also indicated in the right panels.
                We observe a clear separation in the \afe{} plot.
	      }
  \label{fig:salt2_vs_env}
\end{figure*}

% TABLE: linear fits
\begin{table*}
\centering
\caption{Summary of linear fit slopes and their significance for the non-zero slope of SALT2 light-curve fit parameters and Hubble residuals with host galaxy [Fe/H] and [$\alpha$/Fe].}
\label{tab:slope}
\begin{tabular}{l c c c c c}
\hline\hline\\[-1.4em]
				& 	\multicolumn{2}{c}{[Fe/H]} 									&&  \multicolumn{2}{c}{[$\alpha$/Fe]}  \\ \cline{2-3} \cline{5-6} 
				& Birth & Current                             								&& Birth & Current \\
\hline \\[-1em]
$x_{1}$ 			& --0.30 $\pm$ 0.74 (0.4$\sigma$) & --0.80 $\pm$ 0.79 (1.0$\sigma$)	&&  0.91 $\pm$ 0.80 (1.1$\sigma$)	& 0.73 $\pm$ 1.03 (0.7$\sigma$) \\ [0.4em]
$c$       			& --0.08 $\pm$ 0.09 (0.9$\sigma$) & --0.06 $\pm$ 0.10 (0.6$\sigma$)	&& 0.01 $\pm$ 0.10 (0.1$\sigma$) 	& 0.08 $\pm$ 0.12 (0.7$\sigma$) \\ [0.4em]
Hubble residual     	& 0.47 $\pm$ 0.19 (2.4$\sigma$)   &   0.50 $\pm$ 0.21 (2.4$\sigma$)	&& --0.16 $\pm$ 0.23 (0.7$\sigma$)	& --0.37 $\pm$ 0.29 (1.3$\sigma$) \\ [0.4em]
\hline
\end{tabular}
\end{table*}

% TABLE: WM difference.
\begin{table*}
\centering
\caption{The weighted mean of SALT2.4 light-curve fit parameters and Hubble residuals in the different \afe{} environments. We split the sample at \afe{} = 0.35 (cf. \Afe{} = 0.40 based on the \afe{} split value).}
\label{tab:stat}
\begin{tabular}{l  c c c c}
\hline\hline\\[-1.4em]
 & $N_{SN}$  & $x_{1}$  & $c$ & Hubble residual (mag) \\
\hline \\[-0.9em]
\afe{}-rich   & 4  			& $-0.81 \pm 0.31$  				& $0.01 \pm 0.06$     				& $-0.05 \pm 0.10$  \\ [0.4em]
\afe{}-poor & 30  			& $-1.62 \pm 0.10$  				&  $0.00 \pm 0.02$  					&  $0.05 \pm 0.03$   \\ [0.4em]
\hline
Difference   &     			& $0.81 \pm 0.33$  (2.5$\sigma$) & $0.01 \pm 0.06$ (0.2$\sigma$)  & $0.10 \pm 0.10$ (1.0$\sigma$) \\
\hline
\end{tabular}
\end{table*}

The birth environment is more directly related to the SN Ia progenitor star. 
Investigation of the impact of the birth environment on the progenitor star will be of great interest and importance for understanding the underlying physics of SNe Ia.
However, a direct observation and analysis of the progenitor star are difficult to make. 
Thus, here instead, we present the impact of the birth environment on the SN Ia observables, such as SN Ia light-curve shape and colour, and the standardized luminosity, so called the Hubble residual (HR).

Fig.~\ref{fig:salt2_vs_env} shows the results of SALT2.4 light-curve fitter \citep{Guy2007, Guy2010, Betoule2014} as a function of SN Ia progenitor star birth (orange stars) and currently-observed (faint blue) environments.
We note that for Fig.~\ref{fig:salt2_vs_env} we use the 38 (out of 44) SNe Ia that have SALT2.4 light-curve fit parameters, such as $x_{1}$ for the SN Ia light-curve shape and $c$ for the SN Ia colour, and HRs.
Among 38, only 34 normal SNe Ia are used in statistical calculations.
Normal SNe Ia are selected based on the typical cut criteria when making a cosmological sample, such as $|x_{1}| < 3$ and $E\left(B - V\right)_{MW} < 0.15$.
SALT2.4 light-curve fit parameters and HRs are taken from \citet{Kim2019}.
For the HRs calculation, they use $\Omega_{M}$ = 0.3 and $H_0$ = 70 $km \: s^{-1} \: Mpc{-1}$ assuming flat $\Lambda$CDM cosmology.

From linear least-square regression fits (Tab.~\ref{tab:slope}), we find that light-curve fit parameters are insensitive to \feh{} ($<0.9\sigma$), while HRs show a linear trend at 2.4$\sigma$ significance.
For comparison, we also performed the linear fits using \Feh{}.
The \Feh{} trends and the significance with light-curve fit parameters and HRs are very similar to the trends when \feh{} is used.

For the first time, we investigate the relationship between SN Ia properties and [$\alpha$/Fe] in the right panels of Fig.~\ref{fig:salt2_vs_env} and Tab.~\ref{tab:slope} in order to study whether [$\alpha$/Fe] can have an impact on SN Ia properties and split the SN sample into different groups.

Linear least-square regression fits with [$\alpha$/Fe] show that both \afe{} ($<1.1\sigma$) and \Afe{} ($<1.3\sigma$) have no linear trends (see pink lines at the right panels of Fig.~\ref{fig:salt2_vs_env} and Tab.~\ref{tab:slope}).
Interestingly, however, in Fig.~\ref{fig:comp_feh_afe} a clear separation can be observed at \afe{} = 0.35.
We split the sample based on \afe{} = 0.35 (cf. \Afe{} = 0.40 based on \afe{} split value) to investigate a difference in the weighted-mean of each parameter in each environment (Tab.~\ref{tab:stat}).
SNe Ia that exploded in the \afe{}-rich environment have intrinsically wider light-curve shapes than those in the \afe{}-poor environment: $0.81\pm0.33$ (2.5$\sigma$).
The two-sample KS test returns a $p$-value of 0.01 (see orange histograms in the upper-right panel of Fig.~\ref{fig:salt2_vs_env}), meaning they are coming from different populations at a statistically significant level.}
Regarding $c$ and HR, there is no difference in weighted-means in each [$\alpha$/Fe] environment: $0.01 \pm 0.06$ (0.2$\sigma$) for $c$ and $0.10\pm0.10$ mag (1.0$\sigma$) for HR after standard light-curve corrections.
$p$-values of them are 0.67 and 0.27 for $c$ and HR, respectively.
We note here again that the results are based on the small number of SNe Ia, 34 normal SNe Ia, especially only four SNe Ia in the \afe{}-rich group.

%%%
%%%
%		SECTION: Discussion and Conclusion
%%%
%%%
\section{Discussion}
\label{sec:discussion}

In the present work, we trace back 44 SN Ia progenitor star birth environments, such as \feh{} and \afe{}, employing the chemical evolution of galaxies with accurately determined currently-observed stellar population properties of early-type host galaxies, such as stellar population age, \Feh{}, and \Afe{}.

We first present how different the currently-observed and the SN Ia progenitor star birth environments are.
We show that \afe{} has the $30.4^{\text{+10.6}}_{-10.1}\%$ wider range than \Afe{}, while the range of \feh{} is not statistically different to \Feh{} ($17.9^{\text{+26.0}}_{-27.1}\%$).

Considering that the birth environments are more directly linked to the SN Ia progenitor star, we investigate the impact of the birth environment on the SN Ia properties with 34 normal SNe Ia.
Light-curve fit parameters are insensitive to \feh{} ($<0.9\sigma$ for the non-zero slope), while a linear trend is observed with HRs at the 2.4$\sigma$ significance level.
\afe{} has no linear trends with SN Ia light-curve fit parameters and HRs ($<1.1\sigma$).
Interestingly, we can clearly split the SN Ia sample into the two groups based on \afe{} = 0.35: SNe Ia exploded in \afe{}-rich and those in \afe{}-poor groups. 
SNe Ia exploded in different \afe{} groups have different weighted-means of light-curve shape parameters: $0.81\pm0.33$ ($2.5\sigma$).
The $p$-value from the two-sample KS test is 0.01, implying that they are drawn from different populations at a statistically significant level.
In contrast, we do not find any difference in weighted-means ($<1.0\sigma$) and distributions ($p$-value > 0.27) of colour and HRs in different \afe{} groups.
We note that the results are based on the small number of SNe Ia, 34 normal SNe Ia, especially only four SNe Ia in the \afe{}-rich group.

Host galaxy age is thought to be a strong candidate for the origin of the environmental dependence of SNe Ia luminosities \citep[e.g.,][]{Kim2018, Kang2020, Wiseman2023}.
However, in the present work, there is no impact from the host galaxy age, because our samples are considered to be in the birth environment (i.e., 0 Gyr old).
Instead, the host galaxy age effect is absorbed into the birth environment via the galactic chemical evolution, increasing the range of the birth environment.

The present work is performed only with SNe Ia that are exploded in early-type host galaxies, which are mostly older than 2 Gyr old \citep[see][]{Kang2016, Kang2020}.
In this older (passive) environment, SN Ia progenitor stars can have enough stellar evolution time from the main-sequence to the white dwarf stages and then the explosion, whereas in younger (star-forming) environments, evolution time is expected to be very short (order of Myr).
This means that for the latter case, the SN Ia progenitor star birth environment is very similar to the currently-observed status of host galaxies (see the difference of currently-observed and birth [$\alpha$/Fe] in the four [$\alpha$/Fe]-rich samples in Fig.~\ref{fig:comp_feh_afe}).
Thus, in this younger (star-forming) case, it would be suitable to use the currently-observed status of host galaxies.
However, almost half of SNe Ia are observed in older (early-type and passive) environments (e.g., \citealt{Rigault2020} for the low-redshift (z < 0.1) sample, and \citealt{Suzuki2012} for the high-redshift (z > 0.9) sample).
That means that half of the host galaxy properties used in current studies are different from the true SN Ia progenitor star birth environments.
Also, the range of host galaxy properties would be narrower than that of the birth environments.
Because of those two effects, the current environmental dependence studies using the currently-observed status of host galaxy properties could result in a weak correlation between SN Ia properties and host galaxy properties.
In our sample, only 2 out of 34 are shifted from [$\alpha$/Fe]-rich to [$\alpha$/Fe]-poor groups (see right panels of Fig.~\ref{fig:salt2_vs_env}).
This number would have a limited impact on studies of the environmental dependence of SN Ia properties.
However, the difference in [Fe/H] and [$\alpha$/Fe] have an impact when performing the SN Ia explosion modelling work \citep[e.g.,][]{Timmes2003}.
In addition, when we have larger samples, the impact of this group shift is expected to increase.
Therefore, the SN Ia progenitor star birth environments would provide more accurate results in the studies of the environmental dependence of SN Ia luminosities, allowing for a better understanding of the SN Ia progenitor stars.

In order to further confirm and extend our findings, we need more host galaxies, specifically those in younger ([$\alpha$/Fe]-rich) environments.
At the same time, for tracing the SN Ia progenitor star environment, accurate host galaxy properties, such as stellar population age and metallicity, are required. 
Those galaxy properties can be obtained from the spectroscopic observation and absorption line analysis of early-type galaxies.
Obtaining both younger environments and accurate host galaxy properties together from early-type galaxies may seem counter-intuitive, because early-type galaxies are considered to have no star formation.
However, as discussed in Sec.~\ref{sec:intro}, it is known that at least 15 percent of nearby giant early-type galaxies show a sign of recent ($\le$ 1 Gyr) star formation \citep[e.g.,][]{Yi2005, Gomes2016a, Gomes2016b}. 
Given this rate, we require large numbers ($O$($10^3$)) of nearby SNe Ia and their host galaxies to obtain statistically significant results.
Those samples are expected to come from ongoing and planned surveys, like the Zwicky Transient Facility \citep{Bellm2019, Graham2019} and the Rubin Observatory’s Legacy Survey of Space and Time \citep{lsst2009}. 

Given the recent star formation activity found in some early-type galaxies, it is also important to understand the star formation history of each host galaxy. 
From this, we can probe chemical and mass assembly evolution for galaxies to determine the more accurate SN Ia progenitor star birth environment.
Soon, the Time-Domain Extragalactic Survey conducted on the 4-metre Multi-Object Spectrograph Telescope \citep{Swann2019} begins its operation to spectroscopically observe $\sim$21,000 SN Ia host galaxies in 5-year survey time (C. Frohmaier in private communication).
This data will enable us to measure additional host galaxy properties, such as mass and star formation history, by using a full spectrum fitting method.

The present work shows that [$\alpha$/Fe] can clearly distinguish the SN Ia sample into two groups, while [Fe/H] cannot.
As theoretical studies showed the impact of the initial metallicity on the SN Ia luminosity \citep[e.g.,][]{Timmes2003, Kasen2009, Hoflich2010}, it would be an interesting test for the SN Ia explosion model to explain the impact of specifically $\alpha$-elements on the SN Ia explosion.

Combining results from these extended studies would lead us to a better understanding of the origin of the environmental dependence and, consequently, the underlying physics of the SN Ia.
This allows us to utilise SNe Ia as more accurate and precise standard candles for cosmology.

%%% Acknowledgements
\section*{Acknowledgements}

We thank the anonymous referee for constructive suggestions and careful reading to clarify the manuscript. 
Y.-L.K. and I.H. acknowledge support from the Science and Technology Facilities Council [grant number ST/V000713/1]. 
L.G. acknowledges financial support from the Spanish Ministerio de Ciencia e Innovaci\'on (MCIN) and the Agencia Estatal de Investigaci\'on (AEI) 10.13039/501100011033 under the PID2020-115253GA-I00 HOSTFLOWS project, from Centro Superior de Investigaciones Cient\'ificas (CSIC) under the PIE project 20215AT016 and the program Unidad de Excelencia Mar\'ia de Maeztu CEX2020-001058-M, and from the Departament de Recerca i Universitats de la Generalitat de Catalunya through the 2021-SGR-01270 grant.

This analysis used \textsc{\texttt{pandas}}\citep{McKinney2010}, \textsc{\texttt{numpy}}\citep{Harris2020}, \textsc{\texttt{scipy}}\citep{Virtanen2020}, \textsc{\texttt{matplotlib}}\citep{Hunter2007}, and \textsc{\texttt{asymmetric\_uncertainty}} \citep{Gobat2022}.

%For the purpose of open access, the authors has applied a creative commons attribution (CC BY) licence (where permitted by UKRI, ‘open government licence’ or ‘creative commons attribution no-derivatives (CC BY-ND) licence’ may be stated instead) to any author accepted manuscript version arising.

%%%%%%%%%%%%%%%%%%%%%%%%%%%%%%%%%%%%%%%%%%%%%%%%%%
\section*{Data Availability}

All data used in the paper are available on the GitHub webpage: \href{https://github.com/Young-Lo/ProgenitorBirthEnv}{https://github.com/Young-Lo/ProgenitorBirthEnv}.

%%%%%%%%%%%%%%%%%%%% REFERENCES %%%%%%%%%%%%%%%%%%

% The best way to enter references is to use BibTeX:

\bibliographystyle{mnras}
%\bibliography{example} % if your bibtex file is called example.bib

\begin{thebibliography}{99}

% A

% B
\bibitem[\protect\citeauthoryear{Bellm et al.}{2019}]{Bellm2019} Bellm E.~C., Kulkarni S.~R., Graham M.~J., Dekany R., Smith R.~M., Riddle R., Masci F.~J., et al., 2019, PASP, 131, 018002. doi:10.1088/1538-3873/aaecbe
\bibitem[\protect\citeauthoryear{Betoule et al.}{2014}]{Betoule2014} Betoule M., Kessler R., Guy J., Mosher J., Hardin D., Biswas R., Astier P., et al., 2014, A\&A, 568, A22. doi:10.1051/0004-6361/201423413
\bibitem[\protect\citeauthoryear{Brout \& Scolnic}{2021}]{Brout2021} Brout D., Scolnic D., 2021, ApJ, 909, 26. doi:10.3847/1538-4357/abd69b

% C
\bibitem[\protect\citeauthoryear{Childress et al.}{2013}]{Childress2013} Childress M., Aldering G., Antilogus P., Aragon C., Bailey S., Baltay C., Bongard S., et al., 2013, ApJ, 770, 107. doi:10.1088/0004-637X/770/2/107
\bibitem[\protect\citeauthoryear{Childress, Wolf, \& Zahid}{2014}]{Childress2014} Childress M.~J., Wolf C., Zahid H.~J., 2014, MNRAS, 445, 1898. doi:10.1093/mnras/stu1892
\bibitem[\protect\citeauthoryear{Chung et al.}{2013}]{Chung2013} Chung C., Yoon S.-J., Lee S.-Y., Lee Y.-W., 2013, ApJS, 204, 3. doi:10.1088/0067-0049/204/1/3
\bibitem[\protect\citeauthoryear{Chung, Yoon, \& Lee}{2017}]{Chung2017} Chung C., Yoon S.-J., Lee Y.-W., 2017, ApJ, 842, 91. doi:10.3847/1538-4357/aa6f19

% D
\bibitem[\protect\citeauthoryear{Demarque et al.}{2004}]{Demarque2004} Demarque P., Woo J.-H., Kim Y.-C., Yi S.~K., 2004, ApJS, 155, 667. doi:10.1086/424966
\bibitem[\protect\citeauthoryear{de La Rosa et al.}{2011}]{deLaRosa2011} de La Rosa I.~G., La Barbera F., Ferreras I., de Carvalho R.~R., 2011, MNRAS, 418, L74. doi:10.1111/j.1745-3933.2011.01146.x
\bibitem[\protect\citeauthoryear{Dong et al.}{2018}]{Dong2018} Dong H., Olsen K., Lauer T., Saha A., Li Z., Garc{\'\i}a-Benito R., Sch{\"o}del R., 2018, MNRAS, 478, 5379. doi:10.1093/mnras/sty1381

% F

% G
\bibitem[\protect\citeauthoryear{Gobat}{2022}]{Gobat2022} Gobat C., 2022, ascl.soft. ascl:2208.005
\bibitem[\protect\citeauthoryear{Gomes et al.}{2016a}]{Gomes2016a} Gomes J.~M., Papaderos P., V{\'\i}lchez J.~M., Kehrig C., Iglesias-P{\'a}ramo J., Breda I., Lehnert M.~D., et al., 2016, A\&A, 585, A92. doi:10.1051/0004-6361/201525974
\bibitem[\protect\citeauthoryear{Gomes et al.}{2016b}]{Gomes2016b} Gomes J.~M., Papaderos P., Kehrig C., V{\'\i}lchez J.~M., Lehnert M.~D., S{\'a}nchez S.~F., Ziegler B., et al., 2016, A\&A, 588, A68. doi:10.1051/0004-6361/201525976
\bibitem[\protect\citeauthoryear{Graham et al.}{2019}]{Graham2019} Graham M.~J., Kulkarni S.~R., Bellm E.~C., Adams S.~M., Barbarino C., Blagorodnova N., Bodewits D., et al., 2019, PASP, 131, 078001. doi:10.1088/1538-3873/ab006c
\bibitem[\protect\citeauthoryear{Guy et al.}{2007}]{Guy2007} Guy J., Astier P., Baumont S., Hardin D., Pain R., Regnault N., Basa S., et al., 2007, A\&A, 466, 11. doi:10.1051/0004-6361:20066930
\bibitem[\protect\citeauthoryear{Guy et al.}{2010}]{Guy2010} Guy J., Sullivan M., Conley A., Regnault N., Astier P., Balland C., Basa S., et al., 2010, A\&A, 523, A7. doi:10.1051/0004-6361/201014468

% H
\bibitem[\protect\citeauthoryear{Harris et al.}{2020}]{Harris2020} Harris C.~R., Millman K.~J., van der Walt S.~J., Gommers R., Virtanen P., Cournapeau D., Wieser E., et al., 2020, Natur, 585, 357. doi:10.1038/s41586-020-2649-2
\bibitem[\protect\citeauthoryear{H{\"o}flich et al.}{2010}]{Hoflich2010} H{\"o}flich P., Krisciunas K., Khokhlov A.~M., Baron E., Folatelli G., Hamuy M., Phillips M.~M., et al., 2010, ApJ, 710, 444. doi:10.1088/0004-637X/710/1/444
\bibitem[\protect\citeauthoryear{Howell et al.}{2009}]{Howell2009} Howell D.~A., Sullivan M., Brown E.~F., Conley A., Le Borgne D., Hsiao E.~Y., Astier P., et al., 2009, ApJ, 691, 661. doi:10.1088/0004-637X/691/1/661
\bibitem[\protect\citeauthoryear{Hunter}{2007}]{Hunter2007} Hunter J.~D., 2007, CSE, 9, 90. doi:10.1109/MCSE.2007.55

% I

% J
\bibitem[\protect\citeauthoryear{Jha, Riess, \& Kirshner}{2007}]{Jha2007} Jha S., Riess A.~G., Kirshner R.~P., 2007, ApJ, 659, 122. doi:10.1086/512054
\bibitem[\protect\citeauthoryear{Joo \& Lee}{2013}]{Joo2013} Joo S.-J., Lee Y.-W., 2013, ApJ, 762, 36. doi:10.1088/0004-637X/762/1/36

% K
\bibitem[\protect\citeauthoryear{Kang et al.}{2016}]{Kang2016} Kang Y., Kim Y.-L., Lim D., Chung C., Lee Y.-W., 2016, ApJS, 223, 7. doi:10.3847/0067-0049/223/1/7
\bibitem[\protect\citeauthoryear{Kang et al.}{2020}]{Kang2020} Kang Y., Lee Y.-W., Kim Y.-L., Chung C., Ree C.~H., 2020, ApJ, 889, 8. doi:10.3847/1538-4357/ab5afc
\bibitem[\protect\citeauthoryear{Kasen, R{\"o}pke, \& Woosley}{2009}]{Kasen2009} Kasen D., R{\"o}pke F.~K., Woosley S.~E., 2009, Natur, 460, 869. doi:10.1038/nature08256
\bibitem[\protect\citeauthoryear{Kelsey et al.}{2021}]{Kelsey2021} Kelsey L., Sullivan M., Smith M., Wiseman P., Brout D., Davis T.~M., Frohmaier C., et al., 2021, MNRAS, 501, 4861. doi:10.1093/mnras/staa3924
\bibitem[\protect\citeauthoryear{Kelsey et al.}{2023}]{Kelsey2023} Kelsey L., Sullivan M., Wiseman P., Armstrong P., Chen R., Brout D., Davis T.~M., et al., 2023, MNRAS, 519, 3046. doi:10.1093/mnras/stac3711
\bibitem[\protect\citeauthoryear{Kim et al.}{2002}]{Kim2002} Kim Y.-C., Demarque P., Yi S.~K., Alexander D.~R., 2002, ApJS, 143, 499. doi:10.1086/343041
\bibitem[\protect\citeauthoryear{Kim et al.}{2018}]{Kim2018} Kim Y.-L., Smith M., Sullivan M., Lee Y.-W., 2018, ApJ, 854, 24. doi:10.3847/1538-4357/aaa127
\bibitem[\protect\citeauthoryear{Kim, Kang, \& Lee}{2019}]{Kim2019} Kim Y.-L., Kang Y., Lee Y.-W., 2019, JKAS, 52, 181. doi:10.5303/JKAS.2019.52.5.181
\bibitem[\protect\citeauthoryear{Kim et al.}{2024}]{Kim2024} Kim Y.-L., Briday M., Copin Y., Hook I., Rigault M., Smith M., 2024, MNRAS, 527, 4359. doi:10.1093/mnras/stad3501

% L
\bibitem[\protect\citeauthoryear{Lee et al.}{2005}]{Lee2005} Lee Y.-W., Joo S.-J., Han S.-I., Chung C., Ree C.~H., Sohn Y.-J., Kim Y.-C., et al., 2005, ApJL, 621, L57. doi:10.1086/428944
\bibitem[\protect\citeauthoryear{LSST Science Collaboration et al.}{2009}]{lsst2009} LSST Science Collaboration, Abell P.~A., Allison J., Anderson S.~F., Andrew J.~R., Angel J.~R.~P., Armus L., et al., 2009, arXiv, arXiv:0912.0201

% M
\bibitem[\protect\citeauthoryear{Makarov et al.}{2014}]{Makarov2014} Makarov D., Prugniel P., Terekhova N., Courtois H., Vauglin I., 2014, A\&A, 570, A13. doi:10.1051/0004-6361/201423496
\bibitem[\protect\citeauthoryear{Maoz \& Mannucci}{2012}]{Maoz2012} Maoz D., Mannucci F., 2012, PASA, 29, 447. doi:10.1071/AS11052
\bibitem[\protect\citeauthoryear{McKinney}{2010}]{McKinney2010} McKinney W., 2010, in Proceedings of the 9th Python in Science Conference. pp 51–56
\bibitem[\protect\citeauthoryear{Mill{\'a}n-Irigoyen et al.}{2022}]{Millan-Irigoyen2022} Mill{\'a}n-Irigoyen I., del Valle-Espinosa M.~G., Fern{\'a}ndez-Aranda R., Galbany L., Gomes J.~M., Moreno-Raya M., L{\'o}pez-S{\'a}nchez {\'A}. R., et al., 2022, MNRAS, 517, 3312. doi:10.1093/mnras/stac2696

% N

% P
\bibitem[\protect\citeauthoryear{Pan et al.}{2014}]{Pan2014} Pan Y.-C., Sullivan M., Maguire K., Hook I.~M., Nugent P.~E., Howell D.~A., Arcavi I., et al., 2014, MNRAS, 438, 1391. doi:10.1093/mnras/stt2287
\bibitem[\protect\citeauthoryear{Phillips}{1993}]{Phillips1993} Phillips M.~M., 1993, ApJL, 413, L105. doi:10.1086/186970

% R
\bibitem[\protect\citeauthoryear{Rigault et al.}{2013}]{Rigault2013} Rigault M., Copin Y., Aldering G., Antilogus P., Aragon C., Bailey S., Baltay C., et al., 2013, A\&A, 560, A66. doi:10.1051/0004-6361/201322104
\bibitem[\protect\citeauthoryear{Rigault et al.}{2020}]{Rigault2020} Rigault M., Brinnel V., Aldering G., Antilogus P., Aragon C., Bailey S., Baltay C., et al., 2020, A\&A, 644, A176. doi:10.1051/0004-6361/201730404

% S
\bibitem[\protect\citeauthoryear{Saglia et al.}{2018}]{Saglia2018} Saglia R.~P., Opitsch M., Fabricius M.~H., Bender R., Bla{\~n}a M., Gerhard O., 2018, A\&A, 618, A156. doi:10.1051/0004-6361/201732517
\bibitem[\protect\citeauthoryear{Schiavon}{2007}]{S07} Schiavon R.~P., 2007, ApJS, 171, 146. doi:10.1086/511753
\bibitem[\protect\citeauthoryear{Smith et al.}{2020}]{Smith2020} Smith M., Sullivan M., Wiseman P., Kessler R., Scolnic D., Brout D., D'Andrea C.~B., et al., 2020, MNRAS, 494, 4426. doi:10.1093/mnras/staa946
\bibitem[\protect\citeauthoryear{Sullivan et al.}{2010}]{Sullivan2010} Sullivan M., Conley A., Howell D.~A., Neill J.~D., Astier P., Balland C., Basa S., et al., 2010, MNRAS, 406, 782. doi:10.1111/j.1365-2966.2010.16731.x
\bibitem[\protect\citeauthoryear{Suzuki et al.}{2012}]{Suzuki2012} Suzuki N., Rubin D., Lidman C., Aldering G., Amanullah R., Barbary K., Barrientos L.~F., et al., 2012, ApJ, 746, 85. doi:10.1088/0004-637X/746/1/85
\bibitem[\protect\citeauthoryear{Swann et al.}{2019}]{Swann2019} Swann E., Sullivan M., Carrick J., Hoenig S., Hook I., Kotak R., Maguire K., et al., 2019, Msngr, 175, 58. doi:10.18727/0722-6691/5129

% T
\bibitem[\protect\citeauthoryear{Thomas et al.}{2005}]{Thomas2005} Thomas D., Maraston C., Bender R., Mendes de Oliveira C., 2005, ApJ, 621, 673. doi:10.1086/426932
\bibitem[\protect\citeauthoryear{Thomas et al.}{2010}]{Thomas2010} Thomas D., Maraston C., Schawinski K., Sarzi M., Silk J., 2010, MNRAS, 404, 1775. doi:10.1111/j.1365-2966.2010.16427.x
\bibitem[\protect\citeauthoryear{Thomas, Maraston, \& Johansson}{2011}]{TMJ11} Thomas D., Maraston C., Johansson J., 2011, MNRAS, 412, 2183. doi:10.1111/j.1365-2966.2010.18049.x
\bibitem[\protect\citeauthoryear{Timmes, Brown, \& Truran}{2003}]{Timmes2003} Timmes F.~X., Brown E.~F., Truran J.~W., 2003, ApJL, 590, L83. doi:10.1086/376721

% V
\bibitem[\protect\citeauthoryear{Virtanen et al.}{2020}]{Virtanen2020} Virtanen P., Gommers R., Oliphant T.~E., Haberland M., Reddy T., Cournapeau D., Burovski E., et al., 2020, NatMe, 17, 261. doi:10.1038/s41592-019-0686-2

% W
\bibitem[\protect\citeauthoryear{Walcher et al.}{2015}]{Walcher2015} Walcher C.~J., Coelho P.~R.~T., Gallazzi A., Bruzual G., Charlot S., Chiappini C., 2015, A\&A, 582, A46. doi:10.1051/0004-6361/201525924
\bibitem[\protect\citeauthoryear{Williams et al.}{2017}]{Williams2017} Williams B.~F., Dolphin A.~E., Dalcanton J.~J., Weisz D.~R., Bell E.~F., Lewis A.~R., Rosenfield P., et al., 2017, ApJ, 846, 145. doi:10.3847/1538-4357/aa862a
\bibitem[\protect\citeauthoryear{Wiseman et al.}{2023}]{Wiseman2023} Wiseman P., Sullivan M., Smith M., Popovic B., 2023, MNRAS, 520, 6214. doi:10.1093/mnras/stad488


% Y
\bibitem[\protect\citeauthoryear{Yi et al.}{2005}]{Yi2005} Yi S.~K., Yoon S.-J., Kaviraj S., Deharveng J.-M., Rich R.~M., Salim S., Boselli A., et al., 2005, ApJL, 619, L111. doi:10.1086/422811


\end{thebibliography}

% Alternatively you could enter them by hand, like this:
% This method is tedious and prone to error if you have lots of references

%%%%%%%%%%%%%%%%%%%%%%%%%%%%%%%%%%%%%%%%%%%%%%%%%%

%%%%%%%%%%%%%%%%% APPENDICES %%%%%%%%%%%%%%%%%%%%%

\appendix

\section{With different stellar population models}
\label{app:other_models}

\begin{figure}
 \centering
  	\includegraphics[width=0.65\columnwidth]{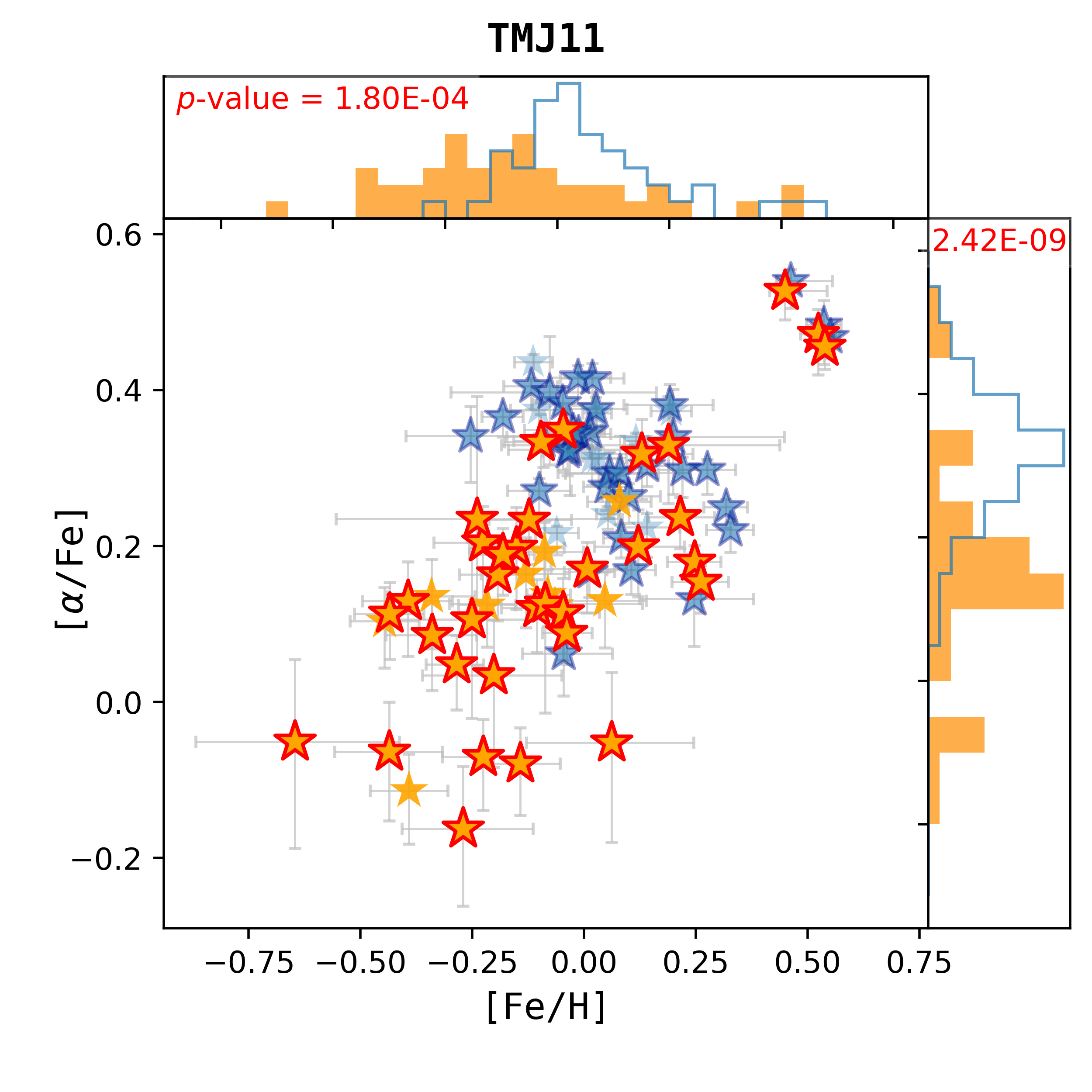}
	\includegraphics[width=\columnwidth]{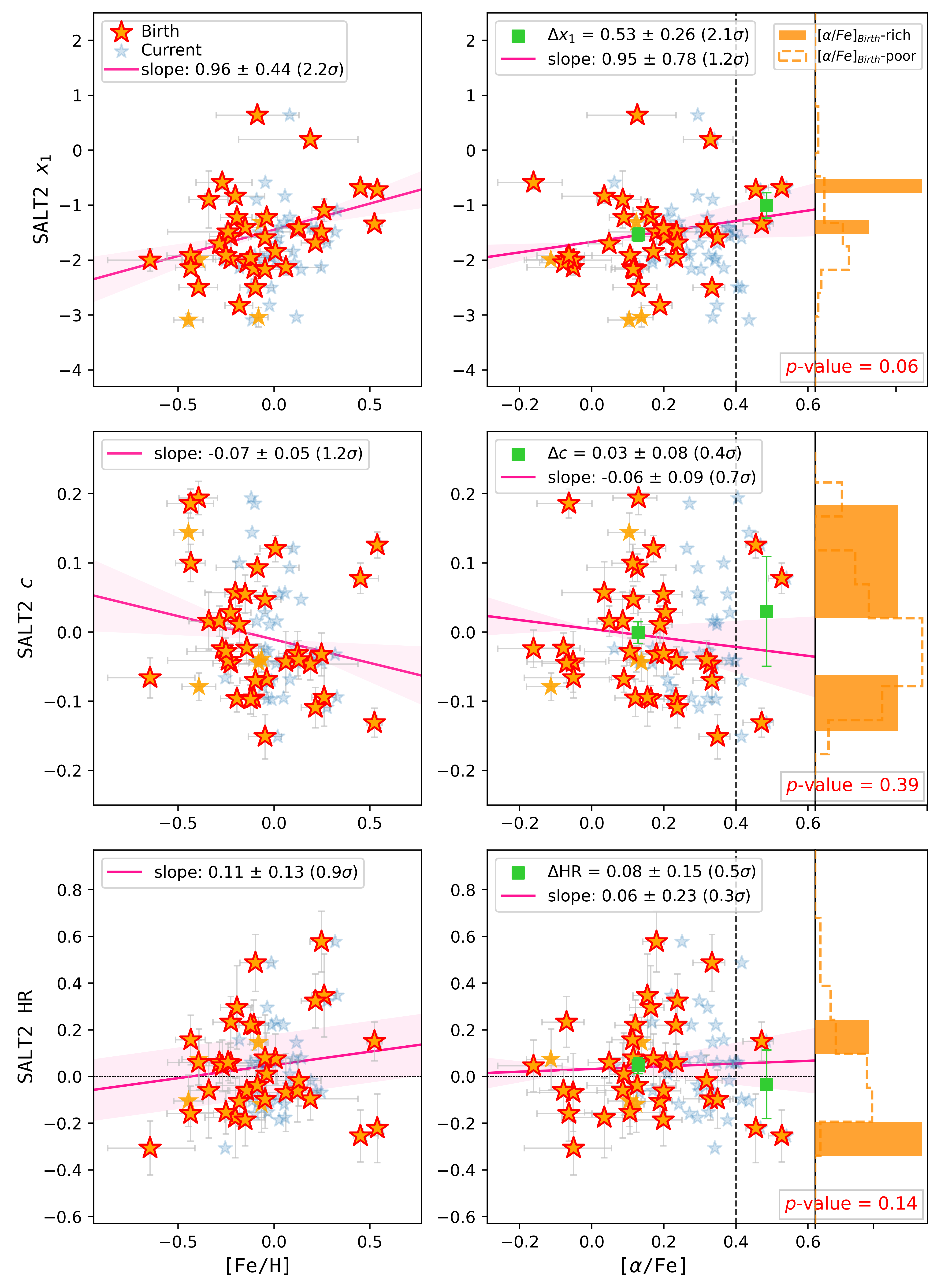}
  \caption{Same as Figs.~\ref{fig:comp_feh_afe} and \ref{fig:salt2_vs_env}, but with the TMJ11 model.
  		[$\alpha$/Fe] environments split at \afe{} = 0.40. 
	      }
  \label{fig:tmj11}
\end{figure}

\begin{figure}
 \centering
  	\includegraphics[width=0.65\columnwidth]{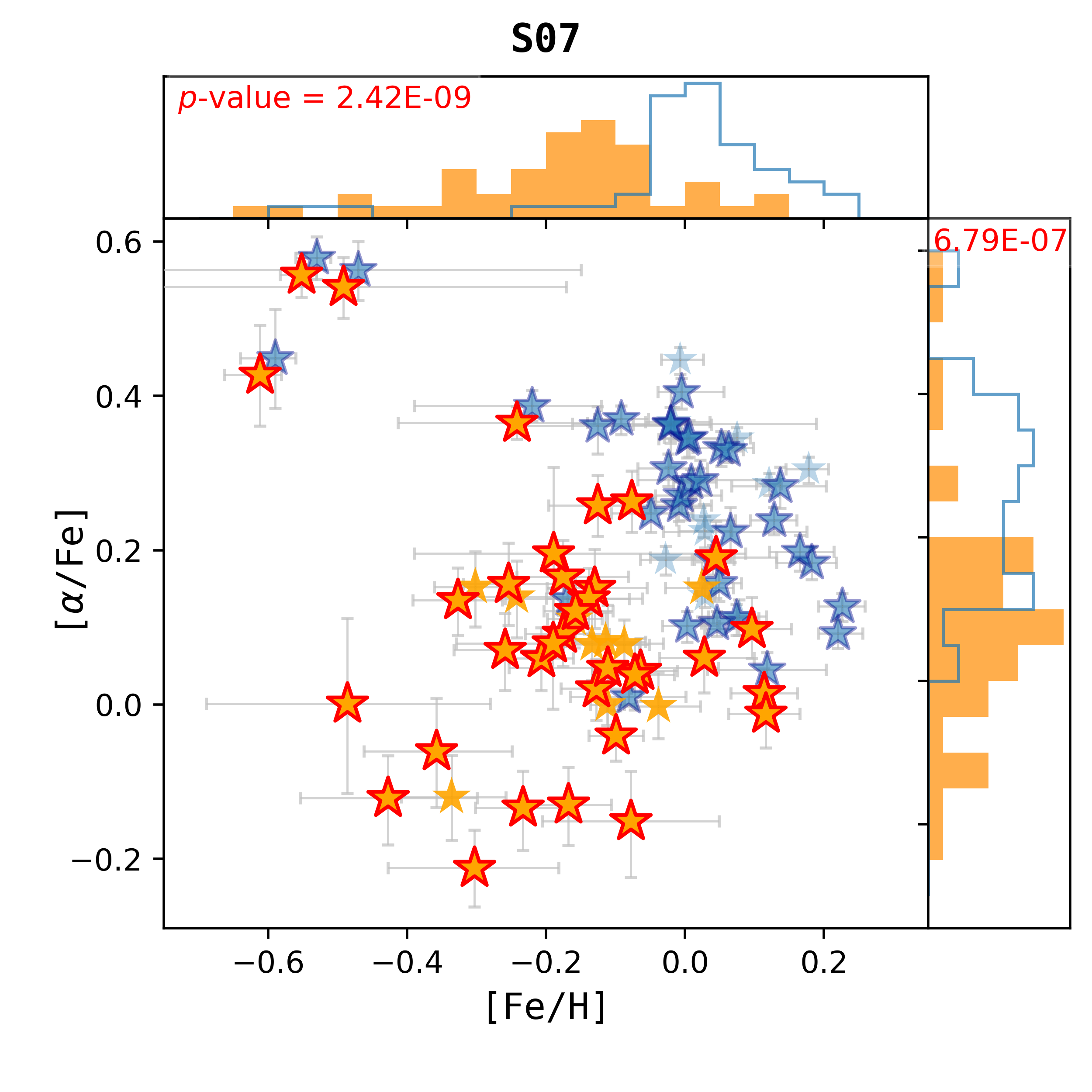}
	\includegraphics[width=\columnwidth]{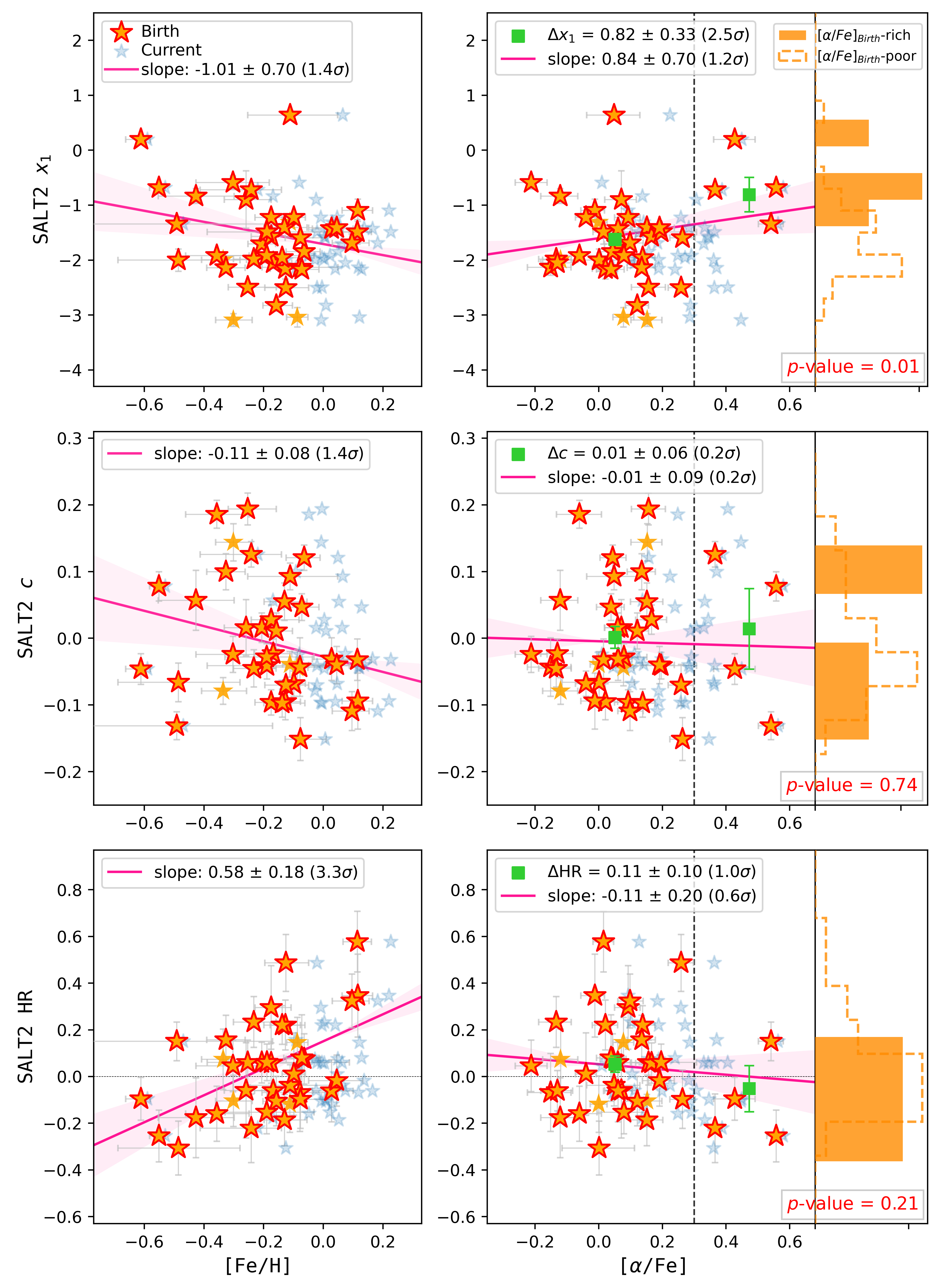}
  \caption{Same as Figs.~\ref{fig:comp_feh_afe} and \ref{fig:salt2_vs_env}, but with the S07 model.
 		 [$\alpha$/Fe] environments split at \afe{} = 0.30. 
	      }
  \label{fig:s07}
\end{figure}

\citet{Kang2020} used three different stellar population models: the Yonsei evolutionary population synthesis model \citep[][hereafter YEPS]{Chung2013}, \citet[][hereafter TMJ11]{TMJ11}, and \citet[][hereafter S07]{S07} models.
Among those, they selected YEPS when they performed the quantitative analysis.
As discussed in \citet{Kang2020}, YEPS is based on the more realistic treatment of Helium burning stage in the stellar evolution modeling and is well-calibrated to the color–magnitude diagrams, integrated colors, and absorption indices of globular clusters in the Milky Way and nearby galaxies \citep{Lee2005, Joo2013, Chung2017}.
It is also producing ages for oldest host galaxies which are not inconsistent with the age of the universe.
Therefore, we also use YEPS as a reference model.
Here, we present the results with other two models: Fig.~\ref{fig:tmj11} for the TMJ11 model and Fig.~\ref{fig:s07} for the S07 model.
The birth [$\alpha$/Fe] environments have $44.2^{\text{+23.8}}_{-22.5}\%$ and $35.2^{\text{+11.5}}_{-11.4}\%$ wider ranges than the currently-observed [$\alpha$/Fe] environments for TMJ11 and S07 models, respectively.
The ranges of birth and currently-observed [Fe/H] are similar: $47.1^{\text{+35.6}}_{-35.2}\%$ for the TMJ11 model and $-10.8^{\text{+10.3}}_{-10.6}\%$ for the S07 model.
However, $p$-values (< 0.001) from the two-sample KS test show that the birth and currently-observed environments are drawn from the different populations.
The dependence of SN Ia properties on the progenitor birth environments qualitatively shows similar trends as we found with YEPS.
Although different stellar population models give different [Fe/H] and [$\alpha$/Fe] values, this demonstrates that our results are independent of the stellar population models.

We note that because a [$\alpha$/Fe] treatment differs by the stellar population model (e.g., see Sec. 2.4 of \citet{Chung2013} for the YEPS model), the location of \afe{} split point is expected to vary across models.
However, a detailed discussion of this is beyond the scope of this paper.
Thus, we selected the location of \afe{} split point for each model based on the \afe{} distribution of each model (upper panels of Fig.s~\ref{fig:tmj11} and \ref{fig:s07}).

% FIGURE: TMJ11.

% FIGURE: S07

%%%%%%%%%%%%%%%%%%%%%%%%%%%%%%%%%%%%%%%%%%%%%%%%%%

% Don't change these lines
\bsp	% typesetting comment
\label{lastpage}
\end{document}